# Spin-Torque Sensors for Energy Efficient High Speed Long Interconnects

Zubair Al Azim, Abhronil Sengupta, Syed Shakib Sarwar, and Kaushik Roy

*Abstract*— **In this paper, we propose a Spin-Torque (ST) based sensing scheme that can enable energy efficient multi-bit long distance interconnect architectures. Current-mode interconnects have recently been proposed to overcome the performance degradations associated with conventional voltage mode Copper (Cu) interconnects. However, the performance of current mode interconnects are limited by analog current sensing transceivers and equalization circuits. As a solution, we propose the use of ST based receivers that use Magnetic Tunnel Junctions (MTJ) and simple digital components for current-to-voltage conversion and do not require analog transceivers. We incorporate Spin-Hall Metal (SHM) in our design to achieve high speed sensing. We show both single and multi-bit operations that reveal major benefits at higher speeds. Our simulation results show that the proposed technique consumes only 3.93-4.72 fJ/bit/mm energy while operating at 1-2 Gbits/sec; which is considerably better than existing charge based interconnects. In addition, Voltage Controlled Magnetic Anisotropy (VCMA) can reduce the required current at the sensor. With the inclusion of VCMA, the energy consumption can be further reduced to 2.02-4.02 fJ/bit/mm.**

## I. INTRODUCTION

With the advancement of CMOS technology towards ultra-scaled dimensions and higher orders of integration, the requirements for long distance interconnects have gone beyond the capabilities of conventional voltage-mode Cu-interconnects [1]. There have been numerous alternate proposals that attempt to resolve this issue and in particular, current-mode links have emerged as one of the most promising solutions for moderately longer lines [2]. Current-mode signaling can facilitate low-voltage interconnection and greatly reduce RC-losses in long lines. However, the current-to-voltage conversion at the receiver for standard CMOS operation leads to performance degradation as power-hungry analog trans-impedance amplifiers are used for this purpose in addition to equalization circuitry [3].

In this work, we present an alternate scheme where current signals are used to switch the magnetization state of a free layer magnet with a Domain-Wall (DW) motion at the receiver. The switching is done by using the DW movement through the action of Spin-Orbit Torque (SOT) generated by Spin-Hall Effect (SHE) [4]. We use a CoFe layer as the free layer magnet with up and down magnetic domains separated by a DW as shown in Fig. 1. This CoFe free layer sits on top of a Spin-Hall Metal (SHM) which is Pt in this case. When a charge current flows through the Pt layer, it exerts a spin polarized current into the CoFe layer as a result of SHE [5][6]. The spin-Hall angle [4], which is the ratio of the produced spin current density to the supplied charge current density, is sufficiently high for this configuration [6]. This spin polarized current exerts a torque on the domain wall (DW), thereby moving the DW in the direction of current flow [5]. This mechanism occurs through the combined action of SHE and Dzyaloshinskii-Moriya Interaction (DMI) (discussed in details in Sec II) [6]. It has been shown experimentally that the DW movement due to SOT generated by SHE and DMI is much faster than the spin transfer torque (STT) action of a spin polarized current passing through the CoFe layer [5]. Moreover, the DW moves even without the application of a magnetic field which makes the process energetically favorable [5]. The magnetization state can be sensed by using MTJs and simple digital CMOS components at the receiver. The paths of current flow for the read and write operations for the MTJ are separate with the use of SHMs, which enhances the reliability of the operation [7]. Additionally, the input current can be adjusted depending on the number of input bits to enable both single and multi-bit operations.

We also propose the use of Voltage Controlled Magnetic Anisotropy (VCMA) [8][9] to further reduce the required input current density that can lead to even more energy efficiency. It has been experimentally demonstrated that an electric field can alter the interfacial anisotropy at the CoFe/MgO interface [10]. As the activation energy barrier for magnetization switching of a magnetic layer is proportional to the anisotropy constant of that layer, it is possible to control the required switching energy by controlling the anisotropy via an electric field [11]. Therefore, with the use of electric field assisted anisotropy reduction, magnetic configuration of a free layer magnet can be altered by much smaller amount of current [12]. However, for the voltage to have a major impact on the overall anisotropy of the magnet, the magnet must be very thin and the overall anisotropy must be dominated by the interfacial contribution [12].

Z. A. Azim, A. Sengupta, S. S. Sarwar, and K. Roy are with the School of Electrical and Computer Engineering, Purdue University, West Lafayette, IN 47907, USA (email: zazim@purdue.edu).

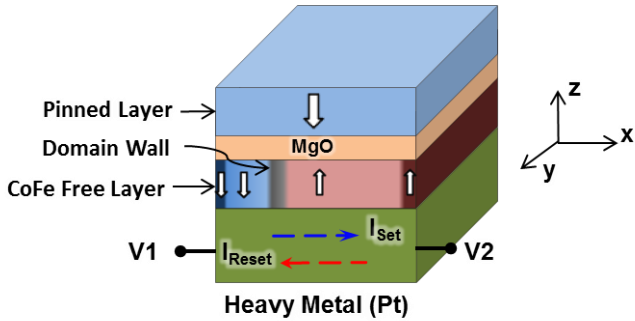

Fig. 1: Domain wall motion (DWM) based device used in the receiver for detecting input current. CoFe layer can be switched in either direction by changing current direction using terminals V1 and V2. The narrow regions on the two edges of the free layer are kept fixed to have a known state for performing the reset operation in circuit implementation.

The CoFe layer used in our proposal is very thin (< 1 nm) (Fig. 1) with predominantly perpendicular magnetic anisotropy (PMA). The PMA in such a small CoFe magnet is dominated by the interface [13]. The rate of interfacial anisotropy change is experimentally found to be in the range of $30 - 50\ \mu J/m^2$ per $V/nm$ for an applied electric field at the CoFe/MgO interface [10][14]. This change in interfacial anisotropy leads to a significant change in the overall anisotropy constant and hence, the energy barrier [11]. Therefore, by applying electric field of desired polarity, the energy barrier of the CoFe free layer can be reduced sufficiently [10]. As a result, the DW can be moved by a smaller input current density through the SHM layer leading to further reduction in energy consumption [15].

Our proposed scheme facilitates signal transmission at ultra-low voltages and avoids the use of analog components at the receiver. The proposed technique is equally applicable to high speed off-chip and global on-chip communication with significant improvement in energy consumption over existing methods as we will demonstrate.

The rest of the article is organized as follows. In section II, we present the details of our proposed device and circuit operation. In section III, we present the method of including VCMA in our design. We present our results and comparison with existing technologies in section IV. Section V concludes the article.

## II. PROPOSED DEVICE AND CIRCUIT OPERATION

### A. Device Operation

We first explain the operation of the device shown in Fig. 1. The ultrathin CoFe layer (0.6 nm thick) lies on top of a heavy metal (3 nm thick Pt). As reported in ref. [16], this configuration demonstrates a strong negative DMI ($DMI\ constant = -1.2\ mJ/m^2$) and a positive spin Hall angle ($\theta_{SH} = 0.07$). Due to the strong DMI, DWs in the CoFe layer get stabilized into Néel type with fixed chirality at rest condition. The chirality is left-handed for this configuration i.e. Down-Right-Up or Up-Left-Down as shown in Fig. 2 [5]. Now, let us consider that a current is flowing in the SHM layer in $+\hat{x}$ direction (electrons flow in $-\hat{x}$ direction). This current induces an effective magnetic field ($\overrightarrow{H_{SHE}}$) which moves the DW. $\overrightarrow{H_{SHE}}$ can be represented as [5]:

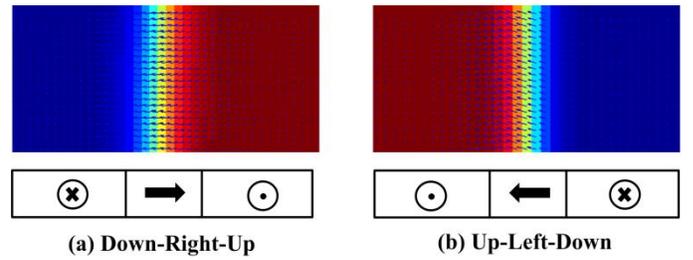

Fig. 2: Left-Handed chirality of Domain Walls for the Pt/CoFe structure.

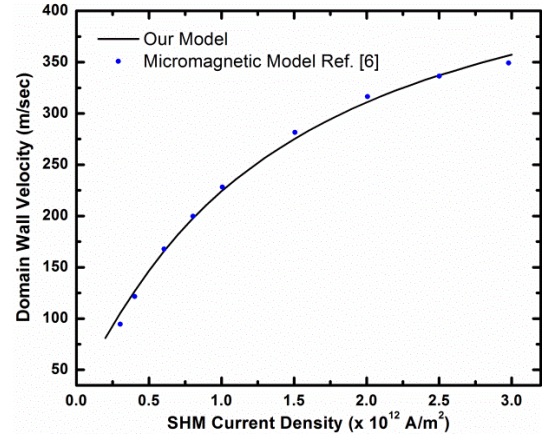

Fig. 3: Benchmarking the proposed simulation framework with Ref. [6] by matching Domain Wall velocity with changing SHM layer current density (No external magnetic field is applied here).

$$\overrightarrow{H_{SHE}} = -\frac{\hbar \theta_{sh} J_q}{2\mu_0 |e| M_s t_f} (\hat{m} \times \hat{\sigma}) \quad (1)$$

where, $\theta_{sh}$ is the spin-hall angle for the structure, $\hat{m}$ is the orientation of the Néel wall, $\hat{\sigma}$ is the direction of injected spins, $t_f$ is the thickness of the SHM layer and $J_q$ is the magnitude of the current flowing through the SHM layer. The direction of this effective field depends on the internal magnetization direction of the Néel DW (Right or Left) and on the magnetization directions of the domains on left and right sides of the DW. This effective field then determines the direction of DW motion [5]. The out-of-plane direction here is $+\hat{z}$ (Fig. 1), so the injected spins are oriented in $+\hat{y}$ direction ($\hat{\sigma} = -\hat{x} \times \hat{z} = \hat{y}$). For the Down-Right-Up DW shown in Fig. 2(a), $\hat{m} = \hat{x}$ and from eqn. (1) we find that $\overrightarrow{H_{SHE}}$ is in $-\hat{z}$ direction which favors the down domain. As a result, DW moves to the right, i.e. along the direction of the current flow. Similarly, for the Up-Left-Down DW shown in Fig. 2(b), $\hat{m} = -\hat{x}$ and from eqn. (1) we find that $\overrightarrow{H_{SHE}}$ is in $+\hat{z}$ direction which favors the up domain. Hence, DW again moves to the right, i.e. along the direction of current flow. Note, the movement of DW is along the current flow regardless of the Up-Down or Down-Up orientation for this structure [5]. The DW can, therefore, be moved in either direction by altering the direction of current flow (using terminals V1 and V2 in Fig. 1). When we later implement this device at the receiver of an interconnect circuit, we need a known initial state for performing a reset operation. Two narrow fixed regions on the edges of the free layer (shown in Fig. 1) are used for this purpose so that the DW moves to a known position next to these narrow fixed regions (during reset operation).

Table I: Design parameters for benchmarking with ref. [6]

| Parameter | Value Used |
|---|---|
| Sat. Magnetization ($M_s$) | 700 kA/m |
| Anisotropy Const. ($K_u$) | $4.8 \times 10^5$ J/m$^3$ |
| Exchange Const. ($A_{ex}$) | $1 \times 10^{-11}$ J/m |
| DMI Const. (D) | -1.2 mJ/m$^2$ |
| Spin Hall angle | 0.07 |
| Gilbert Damping Const. | 0.3 |
| Free layer dimensions | $200 \times 160 \times 0.6$ nm$^3$ |
| SHM dimensions | $200 \times 160 \times 3$ nm$^3$ |
| MgO thickness | 1 nm |

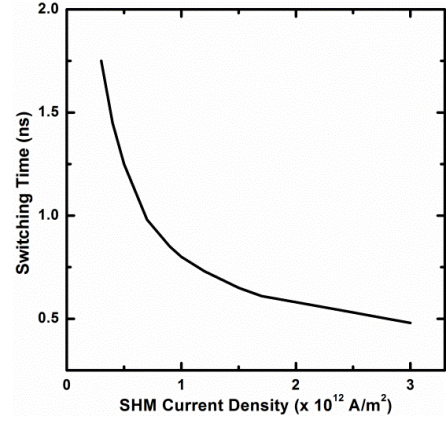

Fig. 5: Time taken to completely switch a 200 nm long CoFe layer with changing current density through the SHM layer

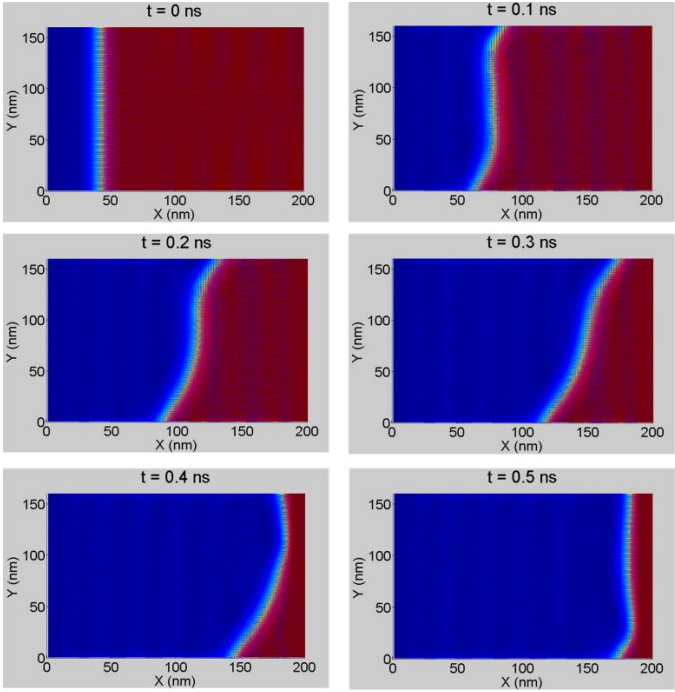

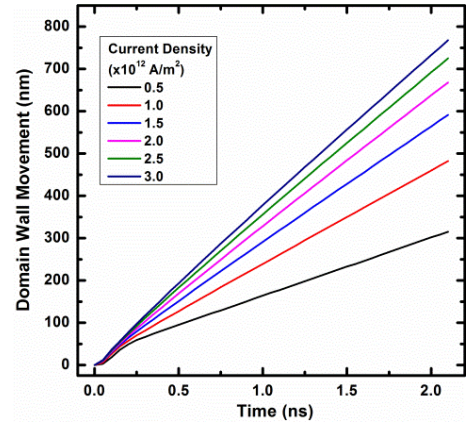

Fig. 6: Domain wall displacements with changing SHE layer current density

Fig. 4: Temporal behavior of Domain Wall movement while current flows through the SHM layer. 'Blue' denotes down-spin and 'Red' denotes up-spin states respectively. DW gets tilted due to the strong DMI.

### B. Modeling and Simulation

In order to characterize the device to circuit operation, we use the mixed mode simulation framework (electron transport, magnetization dynamics from the device to the circuit level) proposed in ref. [17]. The CoFe layer is used as the free layer of an MTJ (shown later in Sec. II(C)), the resistance of which is obtained from the Non-equilibrium Green's Function (NEGF) based simulations [18]. Subsequently, the resistance of the MTJ is used in a MTJ-SPICE model with 45 nm CMOS technology to evaluate the interconnect circuit operations (discussed in Sec. II(C)) [17]. The charge current ($I_e$) flowing through the SHM is obtained from the SPICE simulations and the corresponding spin current ($I_s$) is calculated as [19]:

$$I_s = \theta_{sh} \frac{A_{MTJ}}{A_{SHM}} I_e \sigma \qquad (2)$$

where, $A_{MTJ}$ and $A_{SHM}$ are the cross-sectional areas of the MTJ and the SHM respectively, and $\sigma$ is the polarization of the spin current. The spin current from eqn. (2) is used with generalized LLG equation to analyze the magnetization dynamics as shown below [20]:

$$\frac{d\vec{m}}{dt} = -\gamma_0 \vec{m} \times \overrightarrow{H_{eff}} + \alpha \left( \vec{m} \times \frac{d\vec{m}}{dt} \right) + \frac{1}{qM_s V/\mu_B}(\vec{m} \times \vec{I_s} \times \vec{m}) \qquad (3)$$

where, $\vec{m}$ is the normalized magnetization of the free layer, $\gamma_0$ is the gyromagnetic ratio, α is the Gilbert damping constant, $\overrightarrow{H_{eff}}$ is the effective magnetic field, $M_s$ and $V$ are the saturation magnetization and the volume of the free layer respectively, and $\mu_B$ is the Bohr magneton. We perform these magnetization dynamics simulations using the Mumax3 platform [21]. We first benchmark our simulations by matching the DW velocity against changing SHM layer current density with the micromagnetic simulations from ref. [6] (shown in Fig. 3). The parameters used in the simulations are listed in Table I. DW velocity increases with increasing current density through the SHM layer. The temporal behavior of the DW movement is shown in Fig. 4. The DW gets tilted while moving through the CoFe layer in the direction of current flow due to the presence of strong DMI here [22]. As has been shown experimentally, the DW moves even without applying an external magnetic field [5]. Since the DW velocity increases with current, it takes less time to completely switch a free layer magnet with higher current density through the

SHM layer. This is shown in Fig. 5, where we can observe that the time taken to completely switch a free layer reduces with increasing SHM current density. We also show the different DW displacements in a given time with changing SHM layer current in Fig. 6. As the input current density increases, the DW displacements increase with it in almost linearly. In our proposed circuits, we utilize this feature to detect input current by using the CoFe layer at the receiving end of a long interconnect line. Next, we present the interconnect circuits which are analyzed with the discussed simulation framework.

*C. Circuit Operation*

We now present the interconnect circuits using the Pt/CoFe structure at the receiver. Note, for circuit implementation, we reduce the widths of the CoFe and Pt layers to 20 nm for reducing the required current for DW movement while maintaining the SHM current density as in Fig. 3. The circuit diagram for a single bit interconnect is now shown in Fig. 7. Depending on the data input ($B1\ In$), current is either supplied ($B1\ In = 1$) by the PMOS transistor at the transmitter side. This current first flows through a long Cu-channel (10 mm used for simulations) and then through the SHM layer. The current ($I_{set}$) through the SHM layer induces the DW in the free layer to move to the right when the clock ($CLK$) signal is low (in Fig. 7, when $CLK$ is low, $V_L$ is grounded). When the $CLK$ signal is high, the DW is reset to a known position for the next cycle by passing a reverse current ($I_{reset}$) through the SHM layer (in Fig. 7, when $CLK$ is high, $V_L$ is high). The position of the DW is read using the reference MTJ as shown in the figure which functions as a resistance divider [23]. Standard binary level is detected with a simple CMOS inverter and this is read out ($B1\ Out$) with a positive D-Flip Flop. The normalized magnetization components of the free layer after passing repeated set and reset pulses (S/R) are shown in Fig. 8. The applied S/R pulses are 0.5 ns long ($CLK$ period = 1 ns) which induce altering movements of the DW; thereby, reversing $m_z$ after each alternate pulse. Continuous waveforms at 1 Gbits/sec speed for this scheme are shown in Fig. 9, which shows the operation for three consecutive clock cycles. The node voltages shown in this figure correspond to Fig. 7. Here, during the first two clock cycles, the data input is high ($\overline{B1\ In} = 0$). This results in DW movement to the right and the reversal of free layer magnetization. In the third cycle, the data input is low ($\overline{B1\ In} = 1$) and the magnetization does not reverse. The input pattern is detected at the output through the change in magnetization of the CoFe layer as shown in the waveforms. An ST based interconnect has also been proposed in ref. [23]; however, our proposed technique differs from ref [23] in that, DMI plays a significant role in our device operation and we also incorporate VCMA to boost up the performance (shown later in Sec. III). Moreover, we use a reset based method making it expandable to multi-bit operations. As an example, we show double-bit interconnect operation. For double-bit architecture, the device remains unchanged but the transmitter and detectors are changed as shown in Fig. 10. Here, the two input PMOS transistors are used in the transmitter. These are properly sized to supply four different levels of input currents depending on $B1\ In$ and $B2\ In$ (width of one transistor is set twice that of the other one). Depending on which of the transistors are on, the supplied levels of currents are 0, $I$, $2I$ and $3I$, respectively ($I$ is set to maintain the required current density at the SHM layer for DW movement). For example, if both the transistors are on ($B1\ In = 1$ and $B2\ In = 1$), the amount of supplied current is higher ($3I$) than if only one is on ($2I$ when $B2\ In = 1$, and $I$ when $B1\ In = 1$). With different SHM input current density, the DW displacement is different at the receiver magnet (as shown in Fig. 6); hence the magnetization states are different as well. Reset operation is performed as in the case for single-bit design. For detection, a sample circuit is shown in Fig. 10 where the inverter trip points ($VM1, VM2, VM3$) are properly set to get the desired binary outputs. The output bits ($B1\ Out$ and $B2\ Out$) are again read out using positive D-Flip Flops. The continuous waveforms for four consecutive clock periods ($CLK$ period = 1 ns) are shown in Fig. 11. As shown in the figure, the magnetization states of the free layer are different for different inputs at different clock periods. For example, both input transistors are on during the first period ($\overline{B1\ In} = 0$ and $\overline{B2\ In} = 0$), which results in maximum amount of SHM current supply. As a result, the DW is displaced furthest in this period resulting in the highest change in magnetization as shown in the waveforms. Similarly, for other patterns, the change in magnetization is distinct which results in different outputs. Note, the proposed scheme is applicable for both single and multi-bit operations with only digital components at the receiver. In the next section, we incorporate VCMA in our design which can reduce the required current density for DW movement.

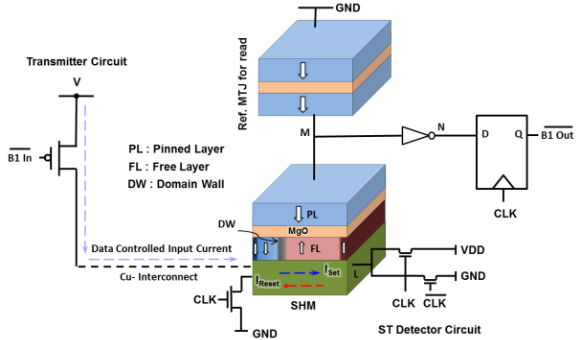

Fig. 7: Single-Bit interconnect architecture with SHM based receiver

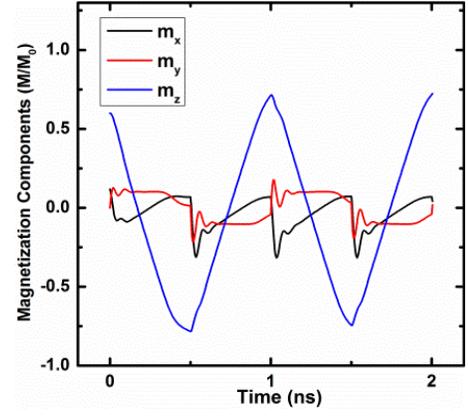

Fig. 8: Free layer normalized magnetization components for repeated switching. $m_z$ reverses after each applied S/R pulse.

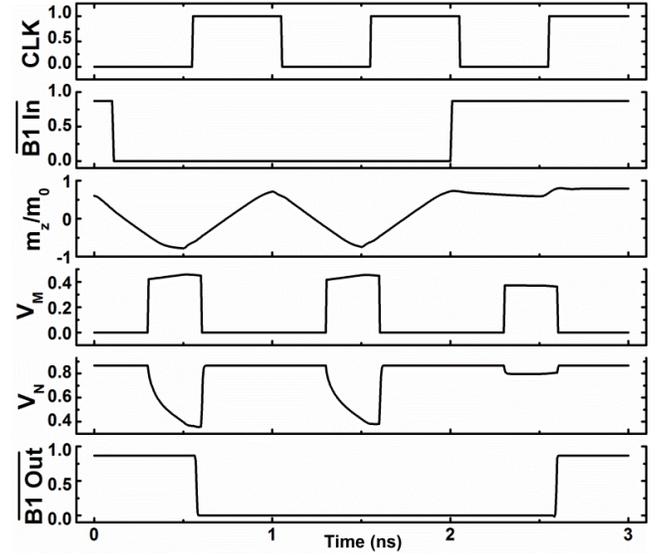

Fig. 9: Simulated waveforms for single-bit transmission (node voltages correspond to Fig. 7)

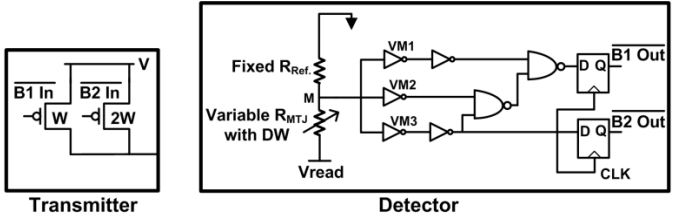

Fig. 10: Transmitter and Detector for double-bit interconnect scheme

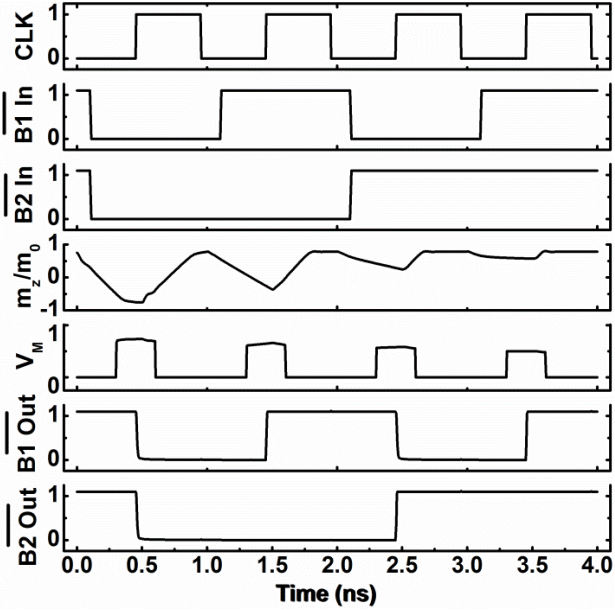

Fig. 11: Simulated waveforms for double-bit transmission (node voltages correspond to Fig. 10)

## III. INCORPORATING VCMA

It has been shown experimentally that an electric field applied at the CoFe/MgO interface can alter the interfacial anisotropy of the CoFe layer [10]. When the MgO is thick enough (~ 1 nm), CoFe/MgO acts as a capacitor and this leads to a voltage induced charge accumulation near the interface between CoFe and MgO layers. If CoFe layer is very thin (< 1 nm), this accumulation of interfacial charge can markedly change the magnetic anisotropy of the CoFe layer. The CoFe free layer proposed in our device is a thin (0.6 nm) perpendicularly polarized layer. In such a thin CoFe layer, PMA is dominated by the interfacial contribution, as the bulk anisotropy is negligible to the interface anisotropy [13]. As a result, when an electric field influences a change in the interfacial anisotropy, this leads to a significant change in the overall anisotropy of the CoFe layer. Theoretically this change in anisotropy occurs due to the change of electron density between d orbital states of the CoFe layer under an applied electric field [24]. Since the energy barrier in a magnet is directly proportional to its anisotropy, it is possible to manipulate the energy barrier using the VCMA effect. Specifically, by using the correct polarity of voltage pulse, the anisotropy of the CoFe layer can be reduced which can lead to lower current requirement to induce the DW movement.

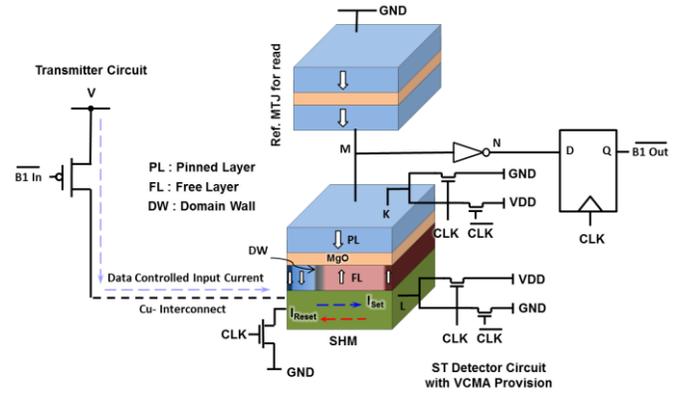

Fig. 12: Single-bit interconnect architecture with VCMA incorporated SHM based receiver

The circuit for including VCMA in our proposed design is shown in Fig. 12. As described earlier, the data dependent DW movement in the CoFe layer occurs during low $CLK$ period. Using the $CLK$ signal, a positive voltage can be applied to the pinned layer with two additional transistors (in Fig. 12 when $CLK$ is low, $V_K$ is high). This voltage is only active when the $CLK$ signal is low, i.e. during the DW movement period. The applied voltage at the top layer leads to an electric field on the order of $1\ V/nm$ at the CoFe/MgO interface ($E = \frac{V}{d} = 1\ V/nm$, $V = 1\ V$, $d = thickness\ of\ MgO = 1 nm$).

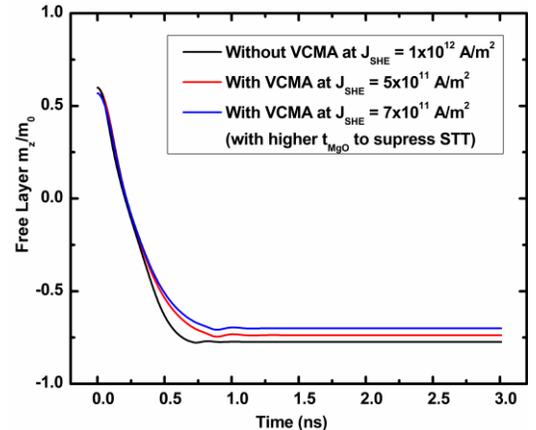

Fig. 13: Comparison of magnetization dynamics with and without VCMA.

With the applied direction, this electric filed induces electron accumulation at the CoFe/MgO interface in the bottom layer. Electron accumulation at the CoFe/MgO interface leads to a decrease of interfacial anisotropy in the CoFe layer [12]. For an applied field of $1\ V/nm$, the reduction of interfacial anisotropy for the CoFe layer is $\sim 38.5\ \mu J/m^2$ (reported experimentally [10][14]). This leads to a considerable reduction of the overall anisotropy of the thin CoFe layer ($\sim 30\%$) [10]. Due to this reduction of anisotropy, the DW in the CoFe layer can be moved with lower current density through the SHM layer [15][25][26]. To observe this effect, we simulated the temporal change of CoFe layer magnetization through DW movement with and without VCMA effect. The applied current density through the SHM layer was kept lower when the VCMA effect was included. As shown in Fig. 13, the temporal responses with and without VCMA effects are comparable even with half the current density through the SHM layer ($5 \times 10^{11}\ A/m^2$ compared to $1 \times 10^{12}\ A/m^2$). The response to a series of Set/Reset pulses with VCMA effect is compared to one without VCMA (Fig. 14). Again, the responses are comparable but with half the SHM current density for the VCMA case, leading to more energy efficiency.

It needs to be mentioned here that, due to the application of a voltage pulse, there will also be a spin transfer torque (STT) on the CoFe layer. To suppress the STT effect, the tunneling current density needs to be reduced which is achievable by increasing the tunnel junction area or by increasing the MgO barrier thickness. These techniques, however, reduces the advantages from the VCMA effect. For instance, we found that increasing the MgO thickness to $1.5\ nm$ keeps the tunneling current density below $10^4\ A/cm^2$. With this tunneling current density, the STT effect is negligible. However, the applied voltage at the top layer now leads to an electric field of $0.67\ V/nm$ at the CoFe/MgO interface ($E = \frac{V}{d} = 0.67\ V/nm, V = 1\ V, d = thickness\ of\ MgO = 1.5\ nm$). With the reduced electric field, the overall anisotropy of the CoFe layer now reduces by $\sim 20\%$. Hence, to observe similar temporal response, the required current density through the SHM layer is now $7 \times 10^{11}\ A/m^2$ (instead of $5 \times 10^{11}\ A/m^2$) as shown in Figs. 13 and 14.

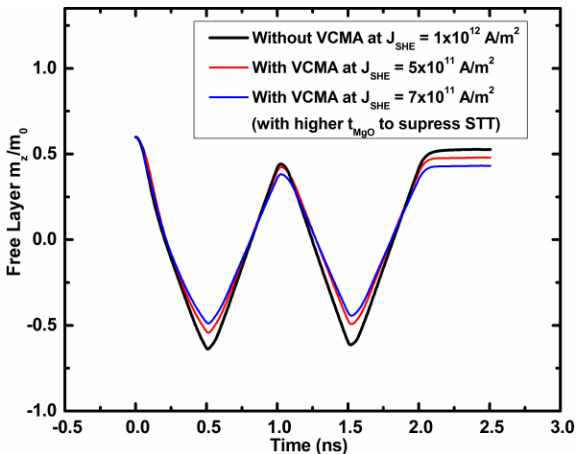

Fig. 14: Free layer normalized magnetization components for repeated switching with VCMA effect. Comparable behavior is observed to without VCMA operation at lesser SHM current density by utilizing VCMA effect.

This SHM current density is still 30% less than without VCMA operation ($1 \times 10^{12}\ A/m^2$). As a result, we still find considerable energy efficiency (shown in next section).

IV. RESULTS AND COMPARISON

The primary advantage of the proposed interconnect is the possibility of low voltage, current mode interconnection without using analog components (amplifiers and equalization circuits) at the receiver. To validate the advantages of the proposed interconnect, we simulated the devices with parameters and operating conditions listed in Table II.

Table II: Device Operating Conditions

| Parameter | Value Used |
|---|---|
| Free layer material | CoFe |
| SHE layer material | Pt |
| Free layer resistivity | 170 μΩ-cm [4] |
| SHM resistivity | 20 μΩ-cm [19] |
| Cu Wire length | 10 mm |
| Wire resistance | 50 Ω/mm [3] |
| Wire capacitance | 0.25 fF/μm [3] |
| SHM current density | $(1-2) \times 10^{12}\ A/m^2$ |
| SHM cross-sectional area | $20 \times 3\ nm^2$ |
| Max. current through SHM | 0.1228 mA |
| Max. voltage across SHM | 82 mV |
| Clock period | 1 ns |
| Data rate for 1-bit design | 1 Gbits/sec |
| Data rate for 2-bit design | 2 Gbits/sec |

Table III: Energy Consumption Analysis

**Single-bit architecture**
$E_{Total} = E_{static} + E_{dynamic} + E_{receiver}$
(1) $E_{static} = E_{static,wire} + E_{static,SHM}$
$r_w = 50\ \Omega/mm, L = 10\ mm, R_{wire} = 500\ \Omega$
$R_{SHM} = 666.67\ \Omega, I_{in} = 0.1228\ mA, T_p = 1\ ns$
$\Rightarrow E_{static,wire} = I_{in}^2 \times R_{wire} \times T_p = 7.5428\ fJ$
$\Rightarrow E_{static,SHM} = E_{SHM,set} + E_{SHM,reset}$
$E_{SHM,set} \rightarrow During\ first\ half\ cycle\ to\ set\ DW$
$E_{SHM,set} = I_{in}^2 \times R_{SHM} \times (T_p/2) = 5.0285\ fJ$
$E_{SHM,reset} \rightarrow During\ second\ half\ cycle\ to\ reset\ DW$
$E_{SHM,reset} = I_{in}^2 \times R_{SHM} \times (T_p/2) = 5.0285\ fJ$
$\Rightarrow E_{static,SHM} = 10.0571\ fJ$
$\Rightarrow E_{static} = 21.1199\ fJ$
(2) $E_{dynamic} \approx CV^2 + E_{driver}$
$c_w = 0.25\ fF/\mu m, L = 10\ mm, E_{driver} = 0.1\ fJ$
$\Rightarrow E_{dynamic} = 16.8618\ fJ$
(3) $E_{receiver} = 4.8717\ fJ\ (SPICE\ analysis)$
Hence, $E_{Total} = 39.3334\ fJ$ for 10 mm wire for 1 bit
$\Rightarrow Energy\ Consumption = \mathbf{3.933\ fJ/bit/mm}$

**Double-bit architecture**
(Similar Analysis) $E_{Total} = 94.4617\ fJ$ for 10 mm for 2 bit
$\Rightarrow Energy\ Consumption = \mathbf{4.7231\ fJ/bit/mm}$

(Similar analysis is performed for including VCMA effect. The results are given in Table IV)

Table IV: Energy consumption comparison with some existing technologies

| Technology | | Energy Consumption (fJ/bit/mm) | Comments |
|---|---|---|---|
| Voltage Mode (Full Swing) | 45 nm CMOS w/o repeaters | 130.743 [27] | Higher delay for longer lines |
| | 45 nm CMOS with repeaters | 302.411 [27] | Lower delay at higher energy |
| Voltage Mode (Low Swing) | 65 nm CMOS w/o repeaters | 8.4-10.9 [28] | Very low speed, requires excessive equalization ckts |
| Current Mode | Sampling Receiver | 35.6 [29] | Large static power in analog current sensing circuits |
| | Sense Amp. Receiver | 9.5 – 10.8 [3] | |
| Optical Interconnects | | ~100 [30] | Very difficult for signal conversion |
| Proposed SHE Based Receiver | Single-bit | 3.93 | Faster and very efficient for multi-bit operation |
| | Double-bit | 4.72 | |
| Proposed SHE Based Receiver with VCMA | Single-bit | 2.02 – 2.53 | Enhanced energy efficiency with the inclusion of VCMA |
| | Double-bit | 3.77 – 4.02 | |

The required current density through the SHM layer for DW movement (without VCMA) is $(1-2) \times 10^{12} \, A/m^2$. This current density results in a maximum required current of 0.1228 mA through the SHM layer and a maximum required voltage of 82 mV across the SHM layer. As a result of such low voltage operation, there is significant reduction in the energy consumption. Table III explains the energy consumption calculations. Total energy consumption consists of the static energy dissipation in the Cu-line and the SHM layer, the dynamic line loss and the energy required for driving the receiver circuit. The calculations in Table III are shown for a 10 mm long Cu-line (parameters used from the experimental implementation in Ref. [3]). The single-bit architecture consumes only 3.93 fJ/bit/mm while the double-bit architecture consumes 4.72 fJ/bit/mm without using VCMA. Note, with VCMA, the current density is reduced by 30-50%. This leads to even lower voltage operation. As a result, there is enhanced energy efficiency. With the inclusion of VCMA and 50% reduction in SHM current density, the single-bit architecture consumes only 2.02 fJ/bit/mm while the double-bit architecture consumes 3.77 fJ/bit/mm. Similarly, for a 30% reduction in SHM current density (with suppressed STT), the single-bit architecture consumes 2.53 fJ/bit/mm and the double-bit architecture consumes 4.02 fJ/bit/mm.

Comparisons of energy consumption with existing charge based conventional interconnects are shown in Table IV. The energy consumption for our proposed method is significantly lower (more than 50x) compared to conventional full-swing voltage mode CMOS interconnects [27]. We also compare the energy consumptions to alternate proposed techniques [3][28][29][30]. In comparison to low-swing voltage mode technology [28], the improvement in energy consumption is ~3x while overcoming the problem of slower operation. As mentioned previously, the major issue in current-mode interconnect is the large static power consumption in analog amplifiers and equalization circuits which is overcome in our proposal. The energy consumption for our technique is ~8x better compared to a sampling receiver based, and ~3x better compared to a sense amplifier receiver based current mode interconnect method [3][29]. Optical interconnect is another promising method for much longer lines [30]. However, for moderately long lines, our method is better both in terms of complexity (difficulty in optical-to-electrical conversion for optical interconnects) and energy consumption (~25x improvement). Finally, the increase in energy consumption at higher bit-rate is negligible for our design which is an excellent feature.

## V. Conclusion

To summarize, we have proposed an interconnect that uses ST sensing to avoid analog transceivers at the receiving end. The proposed design enables fast multi-bit operation by exploiting SHE induced DW movement. The simulation results (calibrated to experiments) show significant energy efficiency over existing techniques in addition to reduced complexity of operation. In addition, the inclusion of VCMA shows further enhancement in performance. The ST-based interconnect can alleviate the issues of existing voltage or current-mode interconnects while achieving enhanced energy efficiency.

Acknowledgment

This research was funded in part by C-SPIN, the center for spintronic materials, interfaces, and architecture, funded by DARPA and MARCO; the Semiconductor Research Corporation, the National Science Foundation, and NSSEFF program.

References

[1] ITRS Roadmap. [Online]. Available: http://www.itrs.net/reports.html, accessed 2013.
[2] N. Tzartzanis, and W. Walker, "Differential current-mode sensing for efficient on-chip global signaling," IEEE Journal of Solid-State Circuits, vol.40, no.11, pp. 2141-2147, Nov. 2005
[3] S. K. Lee, S. H. Lee, D. Sylvester, D. Blaauw and J. Y. Sim, "A 95fJ/b Current-Mode Transceiver for 10mm On-Chip Interconnect," (ISSCC) IEEE International Solid-State Circuits Conference, pp. 262-263, February 2013.
[4] L. Liu, C.-F. Pai, Y. Li, H. W. Tseng, D. C. Ralph, and R. A. Buhrman, "Spin-Torque Switching with the Giant Spin Hall Effect of Tantalum," Science 365, pp. 555-558, 2012.
[5] S. Emori, U. Bauer, S.-M. Ahn, E. Martinez, and G. S. D. Beach, "Current-driven dynamics of chiral ferromagnetic domain walls," Nature Materials 12, pp. 611–616, 2013.
[6] E. Martinez, S. Emori, N. Perez, L. Torres, and G. S. D. Beach, "Current-driven dynamics of Dzyaloshinskii domain walls in the presence of in-plane fields: Full micromagnetic and one-dimensional analysis," JAP 115, 213909 (2014)

[7] Y. Kim, X. Fong, K.-W. Kwon, M.-C. Chen, and K. Roy, "Multilevel Spin-Orbit Torque MRAMs," IEEE Transactions on Electron Devices, Vol. 62, No. 2, February 2015.

[8] M. Weisheit et al., "Electric field-induced modification of magnetism in thin-film ferromagnets," Science, vol. 315, pp. 349-351, 2007.

[9] T. Maruyama et al., "Large voltage-induced magnetic anisotropy change in a few atomic layers of iron," Nat Nano, vol. 4, pp. 158-161, 2009.

[10] Y. Shiota, T. Nozaki, F. Bonell, S. Murakami, T. Shinjo, and Y. Suzuki, "Induction of coherent magnetization switching in a few atomic layers of FeCo using voltage pulses," Nat. Mater., vol. 11, pp. 39-43, 2012.

[11] W. G. Wang, and C. L. Chien, "Voltage-induced switching in magnetic tunnel junctions with perpendicular magnetic anisotropy," J. Phys. D: Appl. Phys. 46, 074004, 2013.

[12] W.-G. Wang, M. Li, S. Hageman, and C. L. Chien, "Electric-field-assisted switching in magnetic tunnel junctions," Nat Mater, vol. 11, pp. 64-68, 2012.

[13] Y. Shiota, T. Maruyama, T. Nozaki1, T. Shinjo, M. Shiraishi, and Y. Suzuki, "oltage-Assisted Magnetization Switching in Ultrathin $Fe_{80}Co_{20}$ Alloy Layers," Applied Physics Express 2 (2009) 063001.

[14] T. Nozaki, Y. Shiota, M. Shiraishi, T. Shinjo, and Y. Suzuki, "Voltage-induced perpendicular magnetic anisotropy change in magnetic tunnel junctions," Applied Physics Letters, vol. 96, p. 022506, 2010.

[15] D. Chiba et al., "Electric-field control of magnetic domain-wall velocity in ultrathin cobalt with perpendicular magnetization," Nature Communications 3, Article number: 888, 2012.

[16] S. Emori et al., "Spin Hall torque magnetometry of Dzyaloshinskii domain walls," Phys. Rev. B 90, 184427 (2014).

[17] X. Fong et al., "KNACK: A hybrid spin-charge mixed-mode simulator for evaluating different genres of spin-transfer torque MRAM bit-cells," in Proc. Int. Conf. SISPAD, Sep. 2011, pp. 51–54.

[18] S. Salahuddin and S. Datta, "Self-consistent simulation of hybrid spintronic devices," in IEDM Tech. Dig., Dec. 2006, pp. 1–4.

[19] S. Manipatruni, D. E. Nikonov, and I. A. Young, "Voltage And Energy-Delay Performance Of Giant Spin Hall Effect Switching For Magnetic Memory And Logic," Cornell Univ. Press, Jan. 2013, pp. 1–16.

[20] B. Behin-Aein, A. Sarkar, and S. Srinivasan, "Switching energy delay of all spin logic devices," Appl. Phys. Lett., vol. 98, no. 12, pp. 123510-1–123510-3, Mar. 2011.

[21] A. Vansteenkiste, J. Leliaert, M. Dvornik, M. Helsen, F. Garcia-Sanchez, and B. V. Waeyenberge, "The design and verification of MuMax3," AIP Advances 4, 107133, 2014.

[22] K.-S. Ryu, L. Thomas, S.-H. Yang, and S. S. P. Parkin, "Current Induced Tilting of Domain Walls in High Velocity Motion along Perpendicularly Magnetized Micron-Sized Co/Ni/Co Racetracks," Appl. Phys. Express 5 093006, 2012.

[23] M. Sharad, X. Fong and K. Roy, "Exploring the Design of Ultra-Low Energy Global Interconnects Based on Spin-Torque Switches," IEDM, pp. 32.6.1-32.6.4, Dec. 2013

[24] K. H. He, J. S. Chen, and Y. P. Feng, "First principles study of the electric field effect on magnetization and magnetic anisotropy of FeCo/MgO(001) thin film," APL 99, 072503 (2011).

[25] U. Bauer, S. Emori, and G. S. D. Beach, "Electric field control of domain wall propagation in Pt/Co/GdOx films," Applied Physics Letters 100, 192408 (2012)

[26] L. Liu, C.-F. Pai, D. C. Ralph, R. A. Buhrman," Gate voltage modulation of spin-Hall-torque-driven magnetic switching," [Online], Available: http://arxiv.org/abs/1209.0962.

[27] Y. Zhang et al.,, "Prediction and Comparison of High-Performance On-Chip Global Interconnection," in Very Large Scale Integration (VLSI) Systems, IEEE Transactions on , vol.19, no.7, pp.1154-1166, July 2011.

[28] J. Postman, P. Chiang, "Energy-efficient transceiver circuits for short-range on-chip interconnects," in Custom Integrated Circuits Conference (CICC), 2011 IEEE , vol., no., pp.1-4, 19-21 Sept. 2011

[29] B. Kim, and V. Stojanovic, "A 4 Gb/s/ch 356 fJ/b 10 mm Equalized On-chip Interconnect with Nonlinear Charge-Injecting Transmit Filter and Transimpedance Receiver in 90 nm CMOS," (ISSCC) IEEE International Solid-State Circuits Conference, pp. 66-67, February 2009.

[30] D. A. B. Miller, "Device requirements for optical interconnects to silicon chips," Proc. IEEE, vol. 97, no. 7, pp. 1166–1185, Jul. 2009.